\documentclass[aps,prb,twocolumn,showpacs,showkeys,floatfix]{revtex4}

\usepackage{graphicx,color}
\graphicspath{{figs/}}
\newcommand{\jwj}[1]{\textcolor{black}{#1}}

\begin{document}
\title{Thermal Conduction in Single-Layer Black Phosphorus: Highly Anisotropic?}
\author{Jin-Wu Jiang}
    \altaffiliation{Email address: jwjiang5918@hotmail.com}
    \affiliation{Shanghai Institute of Applied Mathematics and Mechanics, Shanghai Key Laboratory of Mechanics in Energy Engineering, Shanghai University, Shanghai 200072, People's Republic of China}

%\date{22 December 2009}
\date{\today}
\begin{abstract}

The single-layer black phosphorus is characteristic for its puckered structure, which has leaded to distinct anisotropy in its optical, electronic, and mechanical properties. \jwj{We use the non-equilibrium Green's function approach and the first-principles method to investigate the thermal conductance for single-layer black phosphorus in the ballistic transport regime, in which the phonon-phonon scattering is neglected. We find that the anisotropy in the thermal conduction is very weak for the single-layer black phosphorus -- the difference between two in-plane directions is less than 4\%.} Our phonon calculations disclose that the out-of-plane acoustic phonon branch has lower group velocities in the direction perpendicular to the pucker, as the black phosphorus is softer in this direction, leading to a weakening effect for the thermal conductance in the perpendicular direction. However, the longitudinal acoustic phonon branch behaviors abnormally; i.e., the group velocity of this phonon branch is higher in the perpendicular direction, although the single-layer black phosphorus is softer in this direction. The abnormal behavior of the longitudinal acoustic phonon branch is closely related to the highly anisotropic Poisson's ratio in the single-layer black phosphorus. As a result of the counteraction between the out-of-plane phonon mode and the in-plane phonon modes, the thermal conductance in the perpendicular direction is weaker than the parallel direction, but the anisotropy is pretty small.

\end{abstract}

\maketitle
%\pagebreak

\section{introduction}
Few-layer black phosphorus (BP) is another interesting quasi two-dimensional system that has recently been explored as an alternative electronic material to graphene, boron nitride, and the transition metal dichalcogenides for transistor applications\cite{LiL2014,LiuH2014,BuscemaM2014}. This initial excitement surrounding BP is because unlike graphene, BP has a direct bandgap that is layer-dependent.  Furthermore, BP also exhibits a carrier mobility that is larger than MoS$_{2}$\cite{LiuH2014} and a photo responsivity that is larger than graphene.\cite{LowT2014arxiv}

\begin{figure}[tb]
  \begin{center}
    \scalebox{0.8}[0.8]{\includegraphics[width=8cm]{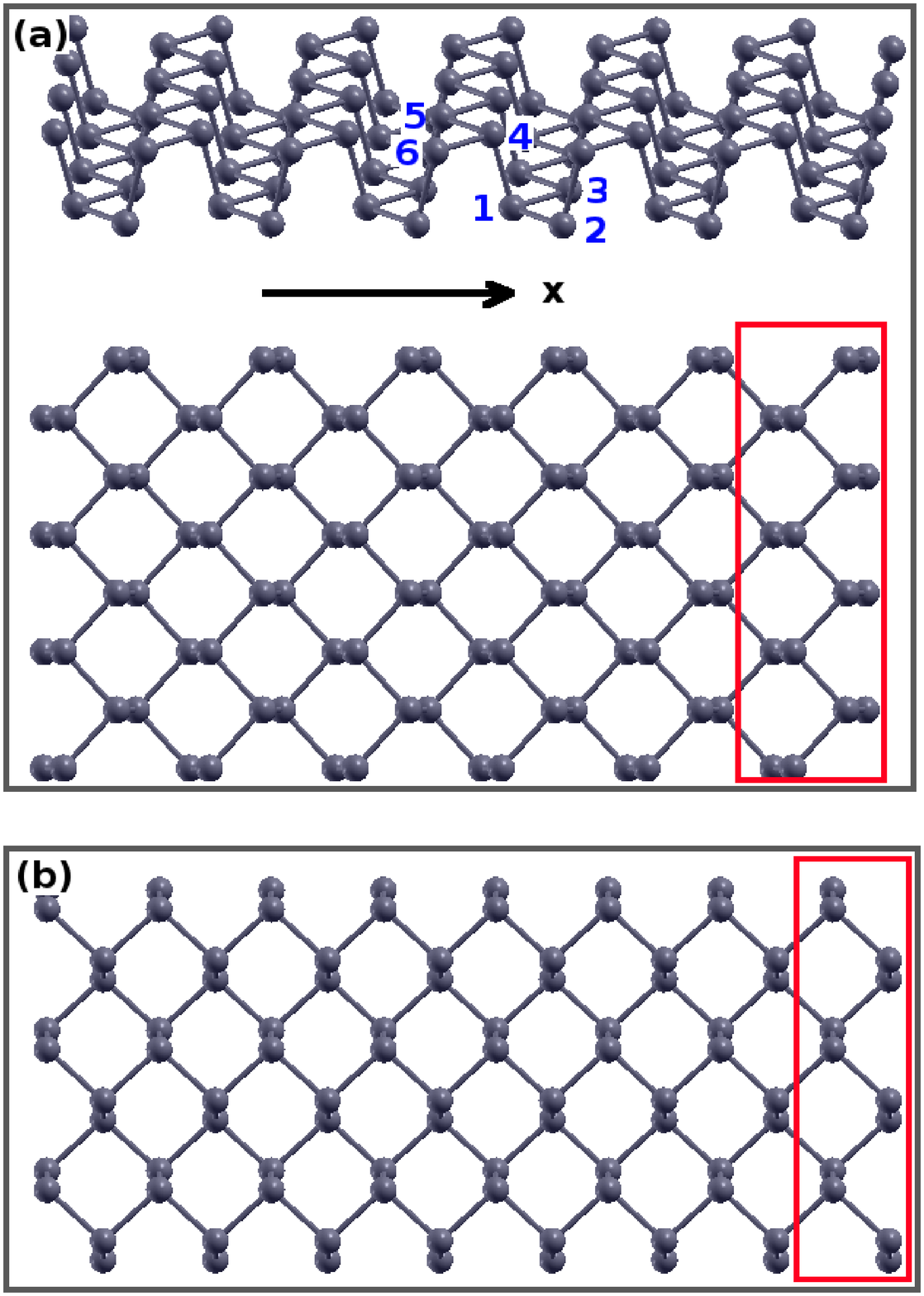}}
  \end{center}
  \caption{(Color online) The relaxed structure of the BPNRs along (a) the perpendicular direction, and (b) parallel direction with respective to the pucker. The $x$-axis is in the direction perpendicular to the pucker. $z$-direction is in the out-of-plane direction. In panel (a), the perspective view illustrates the pucker along the $y$-direction. Red boxes enclose the smallest translational cell in BPNRs.}
  \label{fig_cfg}
\end{figure}

The single-layer BP has a characteristic puckered structure as shown in Fig.~\ref{fig_cfg}~(a), which leads to the two anisotropic in-plane directions. As a result of this puckered configuration, anisotropy has been found in various properties for the single-layer BP, such as the optical properties,\cite{XiaF2014nc,TranV2014prb} the electrical conductance,\cite{FeiR2014nl} and the mechanical properties.\cite{AppalakondaiahS2012prb,QiaoJ2014nc,JiangJW2014bpnpr,QinGarxiv14060261} Furthermore, the Poisson's ratio is so anisotropic that a negative Poisson's ratio was predicted for the single-layer BP, when it is compressed in the direction parallel with the pucker.\cite{JiangJW2014bpnpr,QinGarxiv14060261}

\jwj{It is thus natural to expect that the puckered configuration should also have strong effect on its thermal conduction properties. Indeed, quite recently, Fei et al. found a strong anisotropic thermal transport for single-layer BP in the diffusive transport regime.\cite{FeiR2014nl2} Diffusive transport usually happens at temperatures above room temperature. In the diffusive phonon transport regime, the Fourier law is valid, so the isotropy/anisotropy for the thermal conductivity (a second-order tensor) can be determined by the symmetry of the system.\cite{BornM} If the system has symmetry operation higher than $C_{3}$ rotation, then its thermal conductivity is isotropic. However, for the single-layer BP, the highest-order rotational symmetry is the $C_{2}$ rotational operation, so the thermal conductivity should be anisotropic. Hence, anisotropic thermal conductivity found by Fei et al. is valid according to the fundamental symmetry requirement.}

\jwj{However, in the ballistic phonon transport regime, the Fourier law is invalid, so the above symmetry analysis is not applicable. In this situation, the thermal conductance is purely determined by the phonon dispersion, as the phonon-phonon scattering is not considered in the ballistic regime. The ballistic transport happens at low temperatures in the single-layer BP of high quality, in which the nonlinear phonon-phonon scattering is very weak. So far, it is still unclear whether the ballistic thermal conduction in the single-layer BP is anisotropic.}

\jwj{In this paper, we find that the anisotropy in the thermal conductance is very weak for the single-layer BP in the ballistic transport regime.} The thermal conductance is 0.68~{nWK$^{-1}$} in the direction perpendicular to the pucker, and 0.71~{nWK$^{-1}$} in the parallel direction. Accordingly, the upper limit for the room temperature thermal conductivity in a 1.0~{$\mu$m} long single-layer BP is predicted to be 909.7~{Wm$^{-1}$K$^{-1}$} in the direction perpendicular to the pucker, and 1012.2~{Wm$^{-1}$K$^{-1}$} in the parallel direction. The weak anisotropy in the thermal conductance is discussed based on the phonon dispersion of the single-layer BP.

\section{Structure and simulation details}
In the {\it ab initio} calculation, we use the SIESTA package\cite{SolerJM} to optimize the structure of the single-layer BP. The generalized gradient approximation (GGA) is applied to account for the exchange-correlation function with Perdew-Burke-Ernzerhof (PBE) parametrization\cite{PerdewJP1996prl} and the double-$\zeta$ basis set orbital was adopted. During the conjugate-gradient optimization, the maximum force on each atom is smaller than 0.005 eV\AA$^{-1}$. A mesh cut off of 120 Ry is used. Periodic boundary conditions are applied in the two in-plane transverse directions, while the free boundary condition is applied to the out-of-plane direction by introducing sufficient vacuum space of 15~{\AA}. There are two types of calculations in this work. In the calculation of the phonon dispersion, a single unit cell is used and an $8\times 8\times 1$ Monkhorst-Pack k-grid is chosen for the sampling of the Brillouin zone. In the investigation of the thermal transport, a large translational cell is used and the Gamma point $k$ sampling is adopted.

Fig.~\ref{fig_cfg} shows the relaxed structures for two single-layer BP nanoribbons (BPNRs). In Fig.~\ref{fig_cfg}~(a), the thermal flux is in the direction perpendicular to the pucker, and we will refer to this structure as perpendicular BPNR. In Fig.~\ref{fig_cfg}~(b), the thermal conduction is in the direction parallel with the pucker, which will be referred to parallel BPNR. In the puckered structure, each P atom is connected to three neighboring P atoms. There are two inequivalent P-P bonds in the relaxed structure, i.e $r_{12}=r_{13}=2.4244$~{\AA} and $r_{14}=2.3827$~{\AA}, and also two inequivalent bond angles $\theta_{213}=98.213^{\circ}$ and $\theta_{214}=\theta_{314}=97.640^{\circ}$. These structural parameters are close to the experimental values.\cite{BrownA1965ac} The $x$-axis is along the direction perpendicular to the pucker. The $z$-axis is in the out-of-plane direction.

There are various options to investigate the thermal transport. Classical results can be obtained from the molecular dynamics simulation of the thermal transport, where the phonon-phonon scattering dominates the transport process. From the theoretical point of view, it is a big challenge to provide accurate prediction for the thermal conductivity, because the samples in the experiment always possess various unpreventable defects. Hence, a more practical task is to provide an accurate (quantum) prediction for the upper limit of the thermal conductivity. For this purpose, we would rather apply the ballistic non-equilibrium Green's function (NEGF) method.\cite{WangJSnegf,WangJS2013fp} It is based on quantum mechanics. \jwj{The phonon-phonon scattering is ignored in the ballistic transport region, which is actually quite reasonable for low-dimensional nano-materials at low temperatures. The phonon-phonon scattering is very weak at low temperatures, due to low phonon density. For instance, in graphene, it has been found that the out-of-plane transverse phonon mode can transport almost ballistically even in a large piece of graphene.\cite{NikaDL2009prb}} A proper combination of the NEGF and the {\it ab initio} calculation can provide an accurate upper limit for the thermal conductivity.\cite{JiangJW2011defect}

In the NEGF approach, the thermal conductance is calculated by the Landauer formula:
\begin{eqnarray}
\sigma & = & \frac{1}{2\pi}\int d\omega\hbar\omega T[\omega]\left[\frac{\partial n(\omega,T)}{\partial T}\right],
\label{eq_conductance}
\end{eqnarray}
where $\hbar$ is the Planck's constant. $n(\omega,T)$ is the Bose-Einstein distribution function. The transmission $T[\omega]$ is obtained from the Caroli formula:
\begin{eqnarray}
T[\omega] & = & {\rm Tr}\left(G^{r}\Gamma_{L}G^{a}\Gamma_{R}\right),
\end{eqnarray}
where $G^{r}$ is the retarded Green's function. $G^{a}=\left(G^{r}\right)^{\dagger}$ is the advanced Green's function and $\Gamma_{L/R}$ is the self-energy. These Green's functions can be calculated from the force constant matrix from the {\it ab initio} calculation.\cite{JiangJW2011defect}

\begin{figure}[tb]
  \begin{center}
    \scalebox{1.1}[1.1]{\includegraphics[width=8cm]{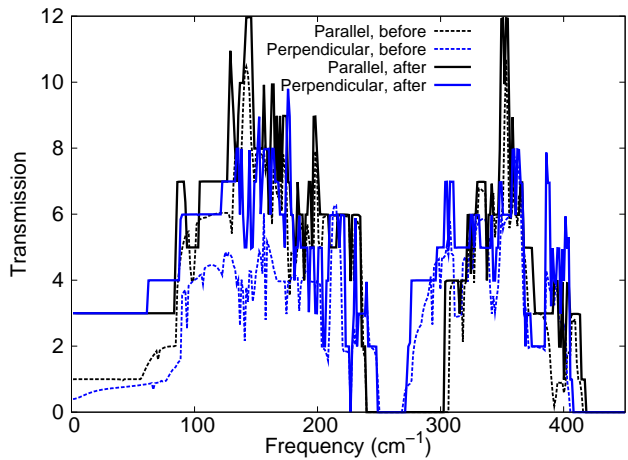}}
  \end{center}
  \caption{(Color online) Transmission functions for the perpendicular and parallel BPNRs. Dashed lines are transmissions before renormalization in the force constant matrix. Solid lines are transmissions after renormalization in the force constant matrix. Note that $T[\omega=0]=3$ after renormalization, which corresponds to the three acoustic phonon branches.}
  \label{fig_transmission}
\end{figure}

\begin{figure}[tb]
  \begin{center}
    \scalebox{1.1}[1.1]{\includegraphics[width=8cm]{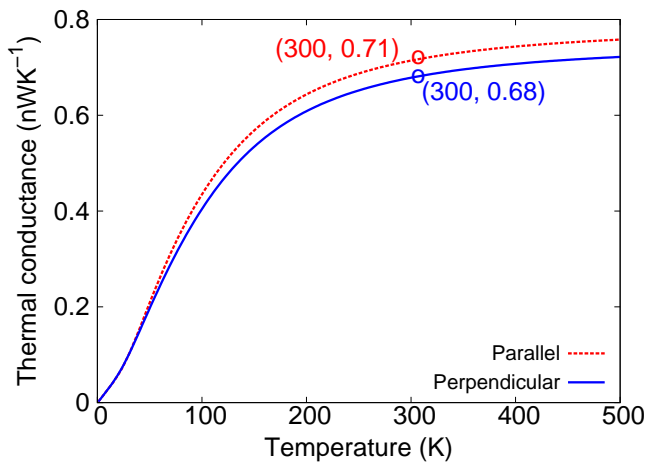}}
  \end{center}
  \caption{(Color online) Thermal conductance for the perpendicular and parallel BPNRs. Two circles enclose the room temperature value. The thermal conductivity can be extracted by $\kappa=\sigma L/s$, where $s$ is the cross-sectional area. The obtained room temperature thermal conductivity is 909.7~{Wm$^{-1}$K$^{-1}$} and 1012.2~{Wm$^{-1}$K$^{-1}$} for the perpendicular and parallel BPNRs, respectively.}
  \label{fig_kappa}
\end{figure}

\begin{figure*}[tb]
  \begin{center}
    \scalebox{0.75}[0.75]{\includegraphics[width=\textwidth]{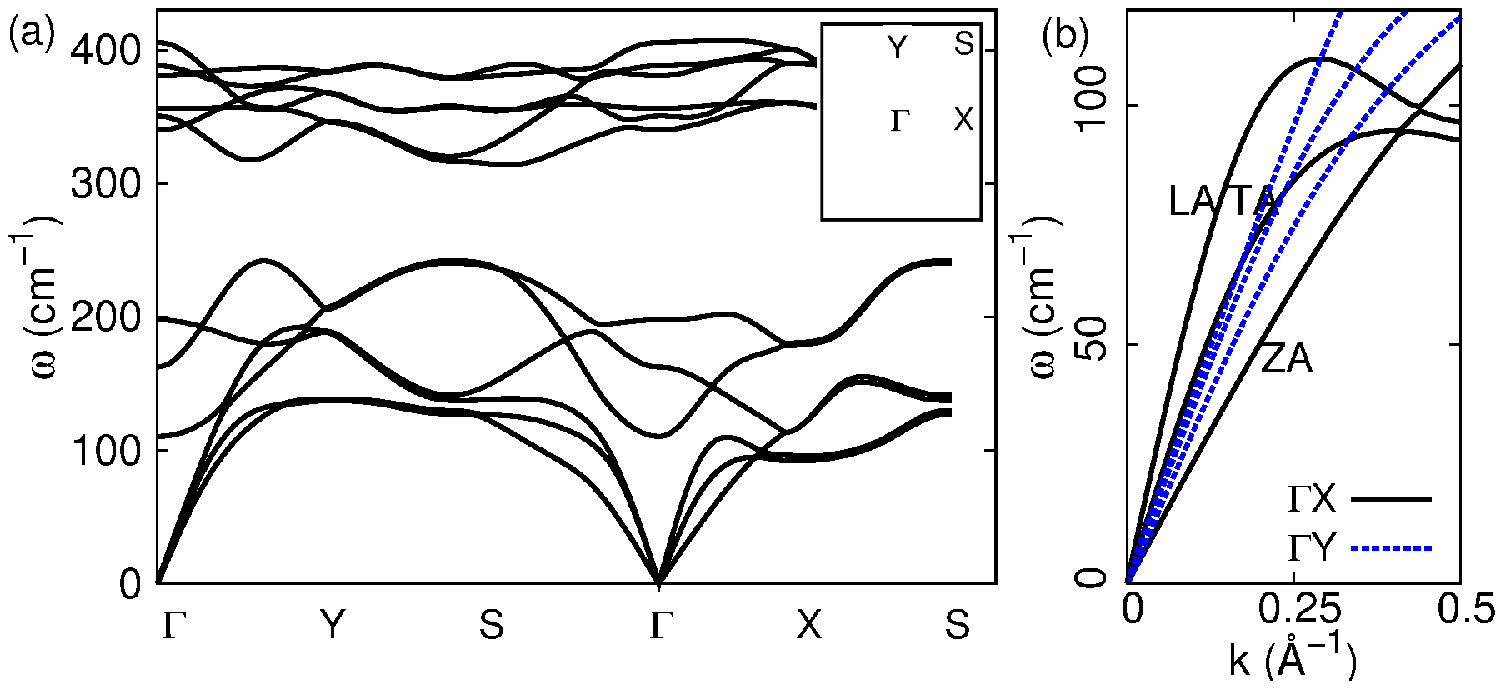}}
  \end{center}
  \caption{(Color online) Phonon dispersion in single-layer BP along high symmetric lines $\Gamma$XS$\Gamma$YS in the Brillouin zone. (a) Full phonon dispersion. Inset shows the first Brillouin zone. $\Gamma$X is along the $x$-axis, i.e., the direction perpendicular to the pucker. (b) The three acoustic branches along $\Gamma$X and $\Gamma$Y directions.}
  \label{fig_phonon}
\end{figure*}

\section{Results and discussions}

The transmission functions are shown in Fig.~\ref{fig_transmission}. As we have discussed in an earlier work, there is a contradiction between the long-range property of the first-principles calculation and the short-range property of the NEGF method.\cite{JiangJW2011defect} We have proposed an efficient renormalization approach to solve this discrepancy. Fig.~\ref{fig_transmission} shows that $T[\omega=0]\not=3$ before renormalization, which is not correct. $T[\omega=0]$ gives the number of acoustic phonon branches, which should be equal to 3 for the single-layer BP. After renormalization, we find that $T[\omega=0]=3$ for both directions. A correct transmission function is very important, as the thermal conductance is calculated based on the transmission function according to Eq.~\ref{eq_conductance}. 

Fig.~\ref{fig_kappa} shows the thermal conductance from the NEGF approach for both perpendicular and parallel BPNRs. It is quite surprising that the thermal conductance is only slightly anisotropic, despite the highly anisotropic puckered configuration of the single-layer BP. The thermal conductance in the perpendicular BPNR is smaller than that of the parallel BPNR in the whole temperature range. At room temperature, the thermal conductance is 0.68~{nWK$^{-1}$} in the perpendicular BPNR, and 0.71~{nWK$^{-1}$} in the parallel BPNR; i.e., the anisotropy in the thermal conductance is only $\xi=|\sigma_{x}-\sigma_{y}|/\sigma_{y} \approx 4\%$. This anisotropy is rather small as compared with the pucker-induced highly anisotropic optical properties,\cite{XiaF2014nc,TranV2014prb} electrical conductance,\cite{FeiR2014nl} and mechanical properties\cite{QiaoJ2014nc,JiangJW2014bpnpr,QinGarxiv14060261}, all of which have anisotropy above 50\%.

The thermal conductance $\sigma$ can be used to get the thermal conductivity ($\kappa$) of a BPNR with arbitrary length $L$: $\kappa=\sigma L/s$, where $s$ is the cross-sectional area. We choose 5.29~{\AA} to be the thickness of single-layer BP as it is the inter-layer spacing in bulk BP.\cite{DuY2010jap} This thickness value is the same for both perpendicular and parallel BPNRs, so its value does not affect our comparison for their thermal conductivity. From the relationship, we get the room temperature thermal conductivity as 909.7~{Wm$^{-1}$K$^{-1}$} in the perpendicular BPNR, and 1012.2~{Wm$^{-1}$K$^{-1}$} in the parallel BPNR. These values are about one fourth of the superior thermal conductivity value (around 5000 Wm$^{-1}$K$^{-1}$) in the graphene,\cite{BalandinAA2008} because the overall phonon spectrum in the BPNR ([0, 400]~cm$^{-1}$) is scaled by a factor of one fourth from the phonon spectrum in the graphene ([0, 1600]~cm$^{-1}$). As we known, the thermal conductivity does not depend on the cross section. It means that the thermal conductivity in BPNRs does not depend on its width, since the thickness is a constant. On the other hand, the thermal conductivity in ballistic regime increases linearly with increasing length, which has been observed in the thermal conductivity of two-dimensional graphene.\cite{BalandinAA2008,NikaDL2009prb,NikaDL2009apl,Balandin2011nm,JiangJW2009direction,XuX2014nc}

The thermal conduction in the ballistic regime is essentially computed based on the phonon dispersion. Hence, we investigate the phonon dispersion of the single-layer BP to examine why the anisotropy in the thermal conductance is so weak. We use the finite difference method to extract the force constant matrix from the first-principles calculations.\cite{YinMT1982prb} Each atom i is displaced in the $\alpha$ (= $\pm x$, $\pm y$, $\pm z$) direction for a small value of $\Delta_{i\alpha}=\Delta=0.04$~{Bohr}, which gives the force constant $C_{i\alpha;j\beta}$,
\begin{eqnarray}
C_{i\alpha;j\beta}=-\frac{F_{i\alpha}}{\Delta_{j\beta}},
\label{eq_c}
\end{eqnarray}
where $F_{i\alpha}$ is the force of atom i in the $\alpha$ direction due to the displacement of atom j in the $\beta$ direction. The force constants in the $x$, $y$, or $z$ direction are obtained from the average of the force constants corresponding to $\pm x$, $\pm y$, or $\pm z$. The dynamical matrix is constructed based on these force constant matrices. The diagonalization of the dynamical matrix results in the phonon frequency and eigenvector, i.e phonon dispersion.

Fig.~\ref{fig_phonon}~(a) shows the phonon dispersion for both acoustic and optical branches in the single-layer BP. There are twelve curves, corresponding to the four P atoms in the primitive unit cell. The three lowest curves correspond to the three acoustic branches; i.e., the $z$-direction acoustic (ZA) mode, in-plane transverse acoustic (TA) mode, and longitudinal acoustic (LA) mode. It should be noted that the ZA branch should be a flexural mode due to the two-dimensional nature of the single-layer BP. This flexural nature disappears in present calculation, owning to the loss of the rigid rotational invariance symmetry in the {\it ab initio} calculation.\cite{JiangJW2006} Inset in Fig.~\ref{fig_phonon}~(a) shows the first Brillouin zone for the single-layer BP. The wave vector on the $\Gamma$X line is along $x$-axis; i.e., in the direction perpendicular to the pucker. It means that phonon modes with wave vectors on the $\Gamma$X line have ``momentum" in the direction perpendicular to the pucker, so these phonon modes contribute to the thermal conductance in the perpendicular BPNR. Similarly, phonon modes on $\Gamma$Y line contribute to the thermal conductance in the parallel BPNR.

Fig.~\ref{fig_phonon}~(b) compares the three acoustic phonon branches along $\Gamma$X and $\Gamma$Y wave vector lines. These three types of acoustic phonon modes have the highest group velocities among all phonon modes, so they make the most important contribution to the thermal conductance in the single-layer BP. We can see obvious anisotropic effect on the phonon dispersion owning to the puckered structure of the single-layer BP. The anisotropic phonon dispersion has also shown up in a recent work.\cite{FeiR2014arxiv} The ZA branch has lower group velocity in the perpendicular direction ($\Gamma$X) than the parallel direction; while the LA mode has higher group velocity in the perpendicular direction. The group velocity ($c$) can be computed through $c=\partial\omega/\partial k$. Using this formula, we find that, in the direction perpendicular to the pucker, the group velocities are 4.77~{kms$^{-1}$}, 7.20~{kms$^{-1}$}, and 9.78~{kms$^{-1}$} for the ZA, TA, and LA phonon modes, respectively. In the direction parallel with the pucker, the group velocities are 5.92~{kms$^{-1}$}, 6.78~{kms$^{-1}$}, and 7.29~{kms$^{-1}$} for the ZA, TA, and LA phonon modes, respectively.

The single-layer BP is softer in the direction perpendicular to the pucker than its parallel direction. More precisely, the single-layer BP has much smaller Young's modulus in the direction perpendicular to the pucker than its parallel direction. The ZA phonon branch has normal behavior in sense that its group velocity are smaller in a softer direction with smaller Young's modulus (in the perpendicular direction). The group velocity of the TA branch is almost the same in these two in-plane directions. However, the LA branch is somehow ``abnormal"; i.e., its group velocity is higher in the softer (perpendicular) direction. To understand this ``abnormal" behavior, we examine quantitatively the mechanical properties in the single-layer BP. Based on the same {\it ab initio} simulation set up as present work, it has been shown that the Young's modulus are $E_{x}=41.3$~{GPa} and $E_{y}=106.4$~{GPa} in the perpendicular and parallel directions, respectively.\cite{JiangJW2014bpyoung} The Poisson's ratio are $\nu_{x}=0.93$ and $\nu_{y}=0.40$ in the perpendicular and parallel directions, respectively.\cite{JiangJW2014bpnpr} The phonon group velocity of the LA mode is proportional to\cite{LandauLD} $\sqrt{E/(1-\nu^2)}$, in which uniform thickness and mass density have been assumed. This formula was developed for isotropic thin plate. However, if we enforce this formula to the anisotropic puckered single-layer BP, we will be able to extract anisotropic phonon group velocities for the perpendicular and parallel directions. We get $c_{x}/c_{y}=\sqrt{E_{x}/E_{y}}\sqrt{(1-\nu_{y}^2)/(1-\nu_{x}^2)}\approx 1.55$. Our calculations in the above show that $c_{x}/c_{y}=9.78/7.29\approx 1.34$, which is comparable with the prediction of the elasticity theorem (1.55). Now it is clear that the abnormal behavior for the LA mode is the result of the highly anisotropic Poisson's ratio in the single-layer BP.

Higher group velocity leads to stronger thermal transport capability, as the phonon mode can transport thermal energy faster. Furthermore, each phonon mode contributes one energy transport channel in the ballistic thermal transport regime. In the above discussions, we have established that, among all three acoustic phonon branches, the ZA phonon branch has smaller group velocity in the direction perpendicular to the pucker. The TA phonon branch has almost the same group velocity in the two in-plane directions. However, the LA phonon branch has larger group velocity in the direction perpendicular to the pucker. As a result, the LA phonon mode helps to enhance the thermal conductance in the direction perpendicular to the pucker, which counteracts with the weakening effect from the ZA phonon modes. Consequently, the anisotropy is very weak in the thermal conductance of the single-layer BP.

\jwj{It should be pointed out that predictions from the present work are applicable in the ballistic transport regime, i.e. at low temperatures without phonon-phonon scattering. The calculated thermal conductance provides an upper limit for future experimental measurements. If the experiment is performed in a diffusive transport regime, then the obtained thermal conduction should be anisotropic as predicted by Fei et al.\cite{FeiR2014nl2}}

\section{conclusion}
We have applied the ballistic NEGF approach to predict the upper thermal conductivity value for the single-layer BP. The force constant matrix is calculated from the first-principles method. Only weak anisotropy is obtained in the thermal conductance of the single-layer BP, despite its highly anisotropic puckered structure. Our phonon dispersion calculations show that while the ZA mode has lower group velocity in the direction perpendicular to the pucker, the LA mode has higher group velocity in the perpendicular direction, which results from highly anisotropic Poisson's ratio in the single-layer BP. The competition between these two opposite effects leads to the weak anisotropic thermal conductance for the single-layer BP.

\textbf{Acknowledgements} The author thanks Xiang-Fan Xu at Tongji University and Jing-Tao L$\rm \ddot{u}$ at HUST for useful communications. The work is supported by the Recruitment Program of Global Youth Experts of China and the start-up funding from Shanghai University.

%\bibliographystyle{aipnum4-1}
%\bibliography{/home/JiangJinWu/Documents/papers/mypapers/latex/biball}
%merlin.mbs aipnum4-1.bst 2010-07-25 4.21a (PWD, AO, DPC) hacked
%Control: key (0)
%Control: author (8) initials jnrlst
%Control: editor formatted (1) identically to author
%Control: production of article title (-1) disabled
%Control: page (0) single
%Control: year (1) truncated
%Control: production of eprint (0) enabled
%

\end{document}